\documentclass[12pt]{article}
\usepackage[margin=2cm]{geometry}
\usepackage{comment}
\usepackage{amsmath,amssymb,extarrows,mathtools,graphicx,subfigure,setspace}
\usepackage{cite}
\usepackage{slashed}
\usepackage{color}
\makeatother

\newcommand{\be}{\begin{equation}}
\newcommand{\bea}{\begin{eqnarray}}
\newcommand{\eea}{\end{eqnarray}}
\newcommand{\ba}{\begin{array}}
\newcommand{\ea}{\end{array}}
\newcommand{\ee}{\end{equation}}
\newcommand{\bes}{\begin{equation*}}
\newcommand{\beas}{\begin{eqnarray*}}
\newcommand{\eeas}{\end{eqnarray*}}
\newcommand{\bas}{\begin{array*}}
\newcommand{\eas}{\end{array*}}
\newcommand{\ees}{\end{equation*}}

\newcommand{\nn}{\nonumber \\ }
\numberwithin{equation}{section}
\begin{document}
\onehalfspacing
\noindent

\begin{titlepage}
\vspace{10mm}
\begin{flushright}
\end{flushright}

\vspace*{20mm}
\begin{center}

{\Large {\bf  On Complexity for $F(R)$ and Critical  Gravity }\\
}

\vspace*{15mm}
\vspace*{1mm}
{Mohsen Alishahiha${}^a$,   Amin Faraji Astaneh$^{b}$,  Ali Naseh${}^b$ and  Mohammad Hassan  Vahidinia${}^{a}$}

 \vspace*{1cm}

{\it ${}^a$ School of Physics, ${}^b$  School of Particles and Accelerators,\\
Institute for Research in Fundamental Sciences (IPM)\\
P.O. Box 19395-5531, Tehran, Iran  }

 \vspace*{0.5cm}
{E-mails: {\tt alishah,faraji,naseh,vahidinia@ipm.ir}}%

\vspace*{1cm}
\end{center}

\begin{abstract}
Using ``complexity=action'' proposal  we study complexity growth of certain gravitational theories
containing higher derivative terms. These include critical gravity in diverse dimensions. 
One observes that the complexity growth for neutral black holes saturates the proposed 
bound when the results are written in terms of physical quantities of the model. 
We will also study effects of shock wave to the complexity growth where we find that 
the  presence of  massive spin-2  mode slows down  the rate of growth.

\end{abstract}

\end{titlepage}

\section{Introduction}

In a gravitational theory a black hole solution may be thought of a thermodynamical system 
whose free energy is given by the finite part of the on-shell action\cite{Gibbons:1997}. 
Although to find the causal structure, symmetries  and Hawking temperature 
of the black hole the explicit form of the solution is enough, for evaluating the entropy, energy
or other charge of the solution the explicit form of the action is needed. This, in turns,
shows the importance of the on-shell action.

Actually the AdS/CFT correspondence\cite{Maldacena:1997re} instructs us to evaluate the bulk action on a given
solution in order to calculate the boundary partition function for a given
boundary metric. This quantity has divergences due to the infinite volume of AdS.
It is then natural to regularize the volume by adding certain
diffeomorphism invariant counter terms which are given in terms of
curvature invariants made out of the induced metric on the boundary
\cite{Henningson:1998gx}. These divergences are typically power law, though in even
boundary dimensions (odd bulk dimension) there are in addition logarithmic
divergent terms  representing the conformal anomaly \cite{Henningson:1998gx}
(see also \cite{Graham:1999pm}).

Using the definition of free energy as on-shell action evaluated on a given solution over the whole 
space time it is natural to introduce the notion of {\it reduced free energy}  as on-shell 
action evaluated over a certain subregion of space time. Indeed, motivated by holographic 
complexity for subregion\cite{Alishahiha:2015rta} (see also \cite{Ben-Ami:2016qex}) a 
distinguished subregion may be given by a
part of space time enclosed by the minimal surface appearing in the computation
of holographic entanglement entropy \cite{Ryu:2006bv}. In particular if one evaluates on-shell action in the volume of an AdS geometry  enclosed by the minimal surface appearing 
in the computation of entanglement entropy for a sphere one finds\cite{Alishahiha}
\be
I_{\rm sphere}=-\frac{\tau}{2\pi \ell} S_{\rm EE},
\ee
where $S_{\rm EE}$ is the holographic entanglement entropy for a sphere with the radius $\ell$ and
$\tau$ is a cut off of the time coordinate.  

More recently in the context of complexity \cite{{Susskind:2014rva},{Stanford:2014jda}} 
it is proposed that on-shell action evaluated on a certain subregion  of the bulk
space time may be related to the complexity of holographic boundary state. 
In this proposal known
as ``complexity=action'' (CA)  
the quantum computational complexity of a holographic state is given by 
the on-shell action evaluated on a bulk region
known as  the ``Wheeler-De Witt'' patch \cite{{Brown:2015bva},{Brown:2015lvg}}
\be
{\cal C}(\Sigma)=\frac{I_{WDW}}{\pi \hbar}.
\ee
Here the Wheeler-De Witt patch (WDW) is defined by domain of dependence of any Cauchy surface in the
bulk whose intersection with the  asymptotically boundary is  the time slice $\Sigma$. It is 
conjectured that there is a bound on complexity growth \cite{Brown:2015lvg}
\be
\frac{d{\cal C}}{dt}\leq \frac{2M}{\pi\hbar},
\ee
that may be thought of as the Lloyd's bound on the quantum complexity\cite{Lloyd:2000}. 
Here $M$ is the mass of  the black hole. For uncharged black holes the bound is saturated. 
It is then important to verify whether the proposed expression for complexity respects the bound.

The main challenge to compute the on-shell action on the  Wheeler-de Witt patch is 
to compute the contribution of boundary terms. Indeed the computation of  the on-shell action
 on the Wheeler-De Witt patch  requires new boundary terms defined on the null boundaries
 as well as on different intersections of
boundaries. For Einstein gravity these boundary terms are found in \cite{Parattu:2015gga}
 ({see also\cite{{Lehner:2016vdi},{Chapman:2016hwi}}), though for a generic gravitational theory 
it is still an open question.

The aim of this paper is to study CA complexity for theories with higher derivative terms. Of course,
as we have already mentioned,  the main challenge is to write the boundary terms. Indeed 
for a generic theory even the corresponding Gibbons-Hawking terms are not known.  
Therefore in this paper we will consider certain higher derivatives terms whose Gibbons-Hawking 
boundary terms on time like boundary are known. More precisely,  we shall  consider two models: a 
gravitational theory whose action is given by a smooth function of 
Ricci scalar ($F(R)$-gravity) and, critical gravity whose action is given by certain 
combination of Ricci and Ricci scalar squared\cite{{Lu:2011zk},{Alishahiha:2011yb}}\footnote{
For Gauss-Bonnet action see \cite{Cai:2016xho}.}.  

Since for these models the corresponding boundary terms on the null boundary are not known,
 in what follows we use the approach of \cite{{Brown:2015bva},{Brown:2015lvg}} to compute
 late time behavior of the complexity growth.  Of course a priori it is not guaranteed that the 
 approach of \cite{{Brown:2015bva},{Brown:2015lvg}}  would also work for  actions with
 higher derivative terms. Nonetheless with the results we have found we are encouraged to 
 follow the procedure for the models under consideration. Indeed in both cases one observes 
 that for uncharged  black hole the complexity growth saturates the bound and therefore 
 respects the Lloyd bound.  
 
The paper is organized as follows. In the next section, we will study the complexity growth for 
 $F(R)$ gravity. In this model we only consider static uncharged black holes. 
 In section three we will compute the complexity growth for static and shock wave solutions
 of critical gravity. The last section is devoted to discussions where we will also explore 
  possibilities of extending ``complexity=volume'' for theories with higher derivative terms.

\section{CA for $F(R$) gravity}

In this section we study  CA proposal of complexity for an $F(R)$ gravity. 
The corresponding  action containing  Gibbons-Hawking term is  given by (see for example \cite{Dyer:2008hb})
\be
I=\frac{1}{16 \pi G}\int_{\mathcal{M}}d^{d+2} x\sqrt{-g} \; F(R)+\frac{1}{8\pi G}\int_{\partial \mathcal{M}}d^{d+1}x\;\sqrt{-\gamma} \; F'(R) K,
\ee
where $F'(R)=\frac{\partial F(R)}{\partial R}$. The equations of motion derived from the above 
action are
\be
F'(R)R_{\mu \nu}-\frac{1}{2} F(R) g_{\mu \nu}-(\nabla_{\mu}\nabla_{\nu}-g_{\mu \nu}\square)F'(R)=0.
\ee
In general it is hard to solve these equations to find a solution, nonetheless  one may look for
solutions satisfying   $R_{\mu \nu}=-\frac{d+1}{\ell^2} g_{\mu \nu}$.
These solutions also include  the AdS-Schwarzschild metric. Indeed plugging this 
ansatz into the equations of motion one finds
\be
F'(R_0)=-\frac{\ell^2}{2(d+1)}F(R_0),\;\;\;\;\;\;\;\;\;\;R_0=-\frac{(d+1)(d+2)}{\ell^2},
\ee
that can be solved to find $\ell$.  Although one could write the corresponding  solution, 
in what follows we do 
not need the explicit form of the metric. Moreover, in general ground,  assuming to have a horizon 
at $r=r_h$ the ADM mass of the corresponding solution is given by \cite{Dyer:2008hb}
\be\label{massF}
M_F=\frac{\Omega_{d}d}{16 \pi G \ell^2}(\ell^2+r_{h}^2)F'(R)r_{h}^{d-1}.
\ee

To evaluate  the on-shell action for the above solution on the WDW patch,  besides the 
generalized Gibbons-Hawking terms given above  one also needs certain boundary terms 
for time like and null  boundaries that in general it is very hard to find them. 
We note, however,  as far as the late time behavior of the complexity growth is concerned 
one may circumvent this major challenge  following the approach considered in \cite{Brown:2015lvg}.
 Actually to follow this procedure we are encouraged by the fact that  for  
 Einstein gravities the results match exactly 
to those obtained using the rigorous method of\cite{Lehner:2016vdi} where all boundary terms have 
been also taken into account. Although a priori it is not clear that this should also work 
for higher derivative terms, as we will see the results we get are consistent with what 
expected. 

To proceed we note that a WDW patch in the bulk is bounded between four light sheets and 
to evaluate the complexity growth one needs two nearby WDW patches and 
the complexity growth is proportional to difference
between on-shell action computed over these two WDW patches.  More precisely,
as it is argued in \cite{Brown:2015lvg}, at late time the only region which lies between two horizons
contributes to this difference (see Fig \ref{fig1}). 
\begin{figure}
\begin{center}
\includegraphics[width=0.65\linewidth]{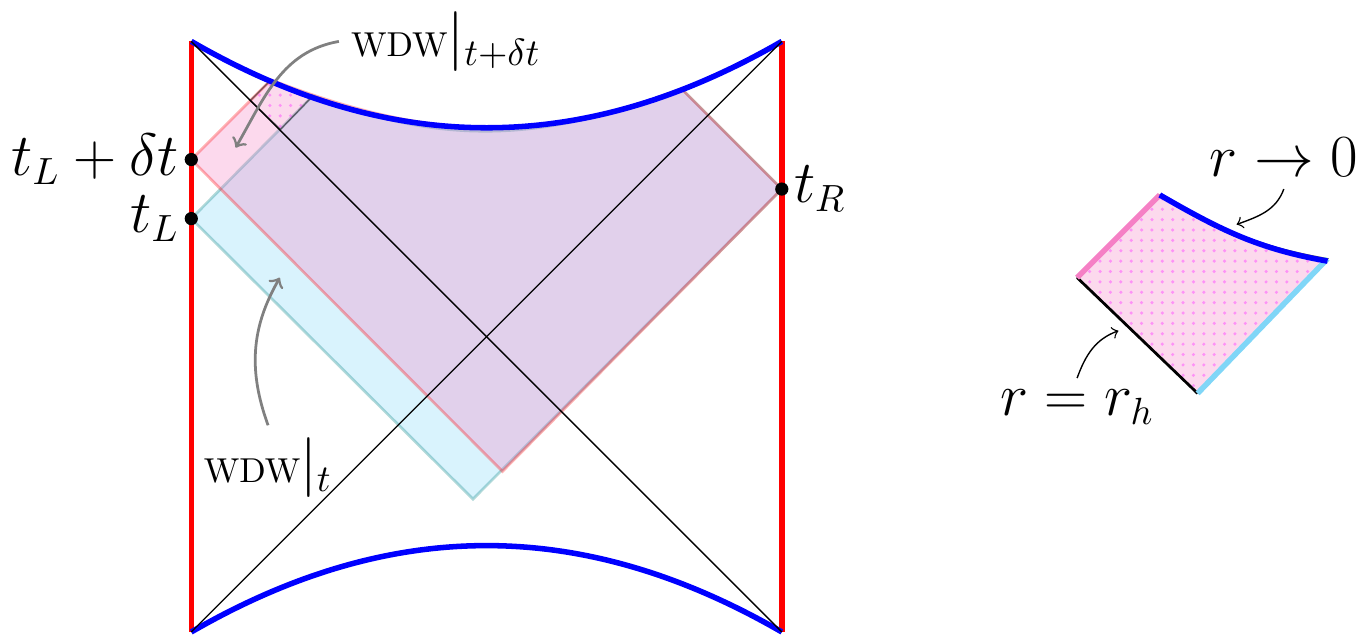}
\caption{Two nearby Wheeler-de Witt patches at $t$ and $t+\delta t$ and the part that contributes to the
late time behavior of complexity growth. This figure is drawn from Fig 2 of \cite{Brown:2015lvg}.}\label{fig1}
\end{center}
\end{figure}
Therefore, we only need to consider the bulk action between
two horizons, {\it i.e.}
\bea
\delta I_{\mathcal{M}} &=& I_{\mathcal{M}} \left[\text{WDW}\vert_{t+\delta t}\right]-I_{\mathcal{M}}
\left[\text{WDW}\vert_{t}\right] \nonumber \\
&=& \frac{1}{16 \pi G} \int_{r \to 0}^{r_{h}}\int_{t}^{t+\delta t}\int_{S^{d}}  \; \mathcal{L}_{F(R)}  \; d
\Omega_{d} \,dt \,dr  \; \ \ \ \text{(at late time)} \nonumber \\
&=& \frac{1}{16 \pi G} \int_{r \to 0}^{r_{h}}\int_{t}^{t+\delta t}\int_{S^{d}} \;r^{d}   \; F(R) \; d
\Omega_{d}\, dt\, dr =\frac{\Omega_{d} F(R)}{16 \pi G (d+1)}r_{h}^{d+1}\, \delta t\,.
\eea
On the other hand from the generalized Gibbons-Hawking term evaluated 
at the horizon and singularity one finds 
\bea
\delta I_{\partial\mathcal{M}} &=& \frac{1}{8 \pi G} \int_{t}^{t+\delta t} \int_{\Omega^{d}} dt\,
d\Omega_{d} \;\left[- \sqrt{f} \; \partial_{r}\left(r^{d}\sqrt{f}\right)F'(R)\right] \Big{\vert}_{r_{h}} \nn
&-&\frac{1}{8 \pi G} \int_{t}^{t+\delta t} \int_{\Omega_{d}} dt\, d\Omega_{d} \;\;\left[- \sqrt{f} \; \partial_{r}\left(r^{d}\sqrt{f}\right)F'(R)\right] \Big{\vert}_{r \to 0} \nn
&=&\frac{\Omega_{d}}{8\pi G}\left(d\,r_{h}^{d-1}+\frac{(d+1)r_{h}^{d+1}}{\ell^{2}}\right)F'(R)\,
\delta t\nn
&=&-\frac{ \Omega_{d} F(R)}{16 \pi G (d+1)}((d+1)r_{h}^2+d\,\ell^2)r_{h}^{d-1}\delta t.
\eea	
Putting these expressions together one arrives at 
\be
\delta(I_{\cal M}+I_{\partial{\cal M}})=-\frac{ \Omega_{d} F(R)d}{16 \pi G (d+1)}(r_{h}^2+\ell^2)
r_{h}^{d-1}\delta t,
\ee
from which one can read the complexity  growth for the eternal AdS-Schwarzschild black hole as
follows
\be
{\dot {\cal C}}=\frac{d I}{dt}=2M_F
\ee
that, interestingly enough, saturates the complexity growth bound. Note that to find the above expression we have used the definition of the ADM mass given by \eqref{massF}.


\section{CA for Critical Gravity}

In this section we will study  the  complexity growth for yet another interesting  theory containing 
higher derivative terms.  The model we will be considering  is ``critical gravity'' whose action is
given by \cite{Lu:2011zk}
 \bea\label{HD}
 I= \frac{1}{16\pi G}\int_{\mathcal{M}} d^{d+2}x \sqrt{-g}\left[R-2\Lambda-\frac{1}{m^2}
  \left(R^{\mu\nu}R_{\mu\nu}-\frac{d+2}{4(d+1)}R^{2}\right)\right],
	\eea
where $m$ is a dimensionful parameter. This model admits several solutions including
AdS and AdS black holes with radius $\ell$. It is known that at the critical point
where  $m^2=\frac{d^2}{2\ell^2}$ the model degenerates yielding to a log-gravity
\cite{Alishahiha:2011yb}.

It is useful to use an auxiliary field $f_{\mu\nu}$ to recast the action \eqref{HD} into the
following form
	\bea
	I=\frac{1}{16\pi G}\int_{\mathcal{M}} d^{d+2}x\sqrt{-g}\left[
	R-2\Lambda+f^{\mu\nu}G_{\mu\nu}+\frac{m^2}{4}\left(f^{\mu\nu}f_{\mu\nu}-f^{2}\right)\right],
	\eea
where $G_{\mu\nu}$ is the Einstein tensor and the auxiliary field $f_{\mu\nu}$ is given by
	\bea
	f_{\mu\nu} = -\frac{2}{m^2}\left(R_{\mu\nu}-\frac{1}{2(d+1)}R g_{\mu\nu}\right).
	\eea	
In this notation the generalized Gibbons-Hawking terms that make
the  variational principle for Dirichlet boundary condition  well posed 
are\cite{Hohm:2010jc}
	\bea\label{GH}
	I_{GGH} = \frac{1}{16\pi G}\int_{\partial\mathcal{M}} d^{d+1}x\sqrt{-\gamma}\left(-2K-\hat{f}^{ij}K_{ij}+\hat{f}K\right).
	\eea
 Here $K_{ij}$ is the extrinsic curvature of the boundary, $K$ is trace of the extrinsic curvature and
 the auxiliary field ${\hat f}^{ij}$ is defined as follows
 \bea
	\hat{f}^{ij} = f^{ij}+2h^{(i}N^{j)}+s N^{i}N^{j},
\eea
where the quantities appearing in this equation are defined via ADM decomposition of
the metric
\be
ds^{2}= N^{2}dr^{2}+\gamma_{ij}(dx^{i}+N^{i}dr)(dx^{j}+N^{j}dr),
\ee
and
\bea
	f^{\mu\nu} = \left(
	\begin{array}{cc}
		s & h^{i} \\
		h^{i} & f^{ij} \\
	\end{array}
	\right)\,.
	\eea
Note also that $\hat{f}=\gamma_{ij}\hat{f}^{ij}$. With this notation in what follows we will compute holographic  complexity for this model using CA  proposal.

\subsection{Stationary solutions}\label{CACG}

In this subsection we would like to investigate the CA proposal of  complexity
 for the critical gravity. As we have already mentioned in order to compute the complexity 
 the  main challenge is to find the contribution of the boundary terms. In what 
 follows we will use the approach of  \cite{Brown:2015lvg} to calculate 
 the late time  growth of complexity.

To start, let us consider three dimensional case
 (known as NMG model\cite{Bergshoeff:2009hq})where one could have a
rotating BTZ black hole whose metric may be given by
	\be\label{BTZ}
	ds^{2}= \frac{dr^{2}}{f(r)}-f(r)dt^{2}+r^{2}\left(d \phi-\frac{8GJ}{2r^{2}}dt\right)^{2},\;\;\;\;\;\;\;\;
	f(r) = \frac{r^{2}}{\ell^{2}}-8GM+\frac{(8GJ)^{2}}{4r^{2}}.	
	\ee
The parameters  $M$ and $ J$ can be expressed in terms  of the outer and inner
 horizon radii, $r_+,r_-$ that
are solutions of the equation $f(r)=0$. More explicitly, one has
\be\label{tt}
f(r)=\frac{(r^{2}-r_{+}^{2})(r^{2}-r_{-}^{2})}{r^{2}\ell^{2}},\hspace{.5cm}8GM=
\frac{r_{+}^{2}+r_{-}^{2}}{\ell^{2}},\hspace{.5cm}8GJ = \frac{2r_{+}r_{-}}{\ell}.
\ee
It is however important to note that  $M$ and $ J$ are the parameters of the solution and
not the physical quantities. Indeed the ADM mass and angular momentum of the solution that
depend on the explicit form of  action are given by\cite{Hohm:2010jc}
\bea\label{yy}
M_{\rm NMG} =M\left(1-\frac{1}{2\ell^{2}m^2}\right),\hspace{.5cm}
J_{\rm NMG} = J\left(1-\frac{1}{2\ell^{2}m^2}\right).
\eea

To compute the complexity growth one needs to consider  difference
between on-shell action evaluated  over two nearby WDW patches depicted in the Fig \ref{fig2}. 
\begin{figure}
\begin{center}
\includegraphics[width=0.42\linewidth]{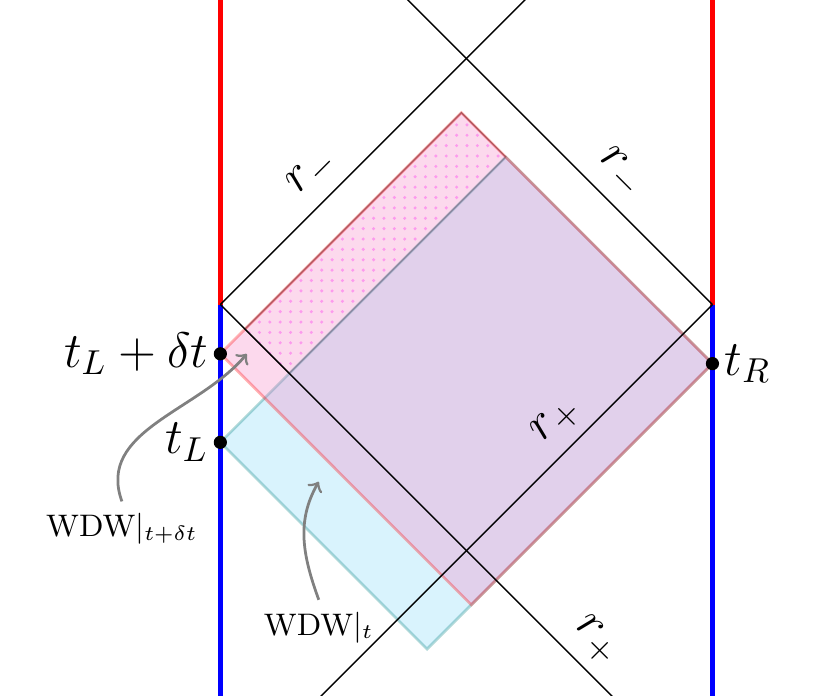}
\caption{Two nearby Wheeler-de Witt patches for BTZ black hole  at 
$t$ and $t+\delta t$ and the part that contributes to the
late time behavior of complexity growth.This figure is drawn from Fig 4 of \cite{Brown:2015lvg}.}
\label{fig2}
\end{center}
\end{figure}
At the late time 
 the only region which lies between two horizons
contributes to this difference. Therefore one gets
 \bea
\delta I_{\mathcal{M}} &=& I_{\mathcal{M}} \left[\text{WDW}\vert_{t+\delta t}\right]-I_{\mathcal{M}}
\left[\text{WDW}\vert_{t}\right] \nonumber \\
&=& \frac{1}{16 \pi G} \int_{r_{-}}^{r_{+}}\int_{t}^{t+\delta t}\int_{0}^{2 \pi} \; \mathcal{L}_{\rm NMG}
 \; d\phi\, dt\, dr  \; \ \ \ \text{(at late time)}\nonumber \\
&=&- \frac{2 \pi \delta t}{16 \pi G}\int_{r_{-}}^{r_{+}}\frac{4 r}{\ell^2}
\left(1-\frac{1}{2 \ell^2m^2}\right)dr
=  -\frac{(r_{+}^{2}-r_{-}^{2})}{4G\ell^{2}}
\left(1-\frac{1}{2\ell^{2}m^2}\right) \delta t.
\eea
The contribution of the generalized Gibbons-Hawking terms  \eqref{GH} are also given by
\bea
\delta I_{\partial\mathcal{M}} &=& \frac{1}{16 \pi G} \int_{t}^{t+\delta t} \int_{0}^{2\pi} dt\, d\phi \;
\left[\frac{-2}{\ell^2}(r_{+}^2+r_{-}^2-2r^2) \left(1-\frac{1}{2 \ell^2m^2}\right)\right] \Big{\vert}_{r_{+}}
\nn &&-\frac{1}{16 \pi G} \int_{t}^{t+\delta t} \int_{0}^{2\pi} dt\, d\phi \;\left[\frac{-2}{\ell^2}
(r_{+}^2+r_{-}^2-2r^2) \left(1-\frac{1}{2 \ell^2m^2}\right)\right] \Big{\vert}_{r_{-}} \nn
&=&\frac{(r_{+}^{2}-r_{-}^{2})}{2G\ell^{2}}
\left(1-\frac{1}{2\ell^{2}m^2}\right) {\delta t}.
\eea	
Therefore,  one arrives at
\bea\label{rateNMG}
\frac{d I}{dt}\equiv\frac{d I_{\mathcal{M}}}{dt}+\frac{d I_{\partial\mathcal{M}}}{dt}=\frac{(r_{+}^{2}-
r_{-}^{2})}{4G\ell^{2}}\left(1-\frac{1}{2\ell^{2}m^2}\right),
\eea
which  is the total rate of growth of the action. By making use of equations \eqref{tt} and \eqref{yy}
the above expression reads
\bea
{\dot {\cal C}}=\frac{d I}{dt} = 2\sqrt{M^{2}-\frac{J^{2}}{\ell^{2}}}\left(1-\frac{1}{2\ell^{2}m^2}\right)=
2\sqrt{M_{\rm NMG}^{2}-\frac{J_{\rm NMG}^{2}}{\ell^{2}}},
\eea
thats is  consistent with results of \cite{Brown:2015lvg}.

Encouraged  by this result let us consider higher dimensional critical gravity. It is
easy to see that the equations of motion obtained from the action \eqref{HD}
admits the AdS-Schwarzschild black hole whose metric is given by
\be\label{metric}
ds^{2} =\frac{dr^{2}}{f(r)}-f(r)dt^{2}+r^{2}d\Omega^{2}_{d},\;\;\;\;\;\;\;\;\;\;
f(r) = \frac{r^{2}}{\ell^{2}}+1-\frac{8\pi}{d\,\Omega_{d}}\frac{2GM}{r^{d-1}}.
\ee
where $M$ is a parameter of the solution (not the physical mass) and the radius $\ell$ satisfies
the following equation
\be
\frac{2\Lambda}{d(d+1)}\ell^4+\ell^2-\frac{d(d-2)}{4m^2}=0.
\ee
Once again we would like to calculate the on-shell action over difference of two WDW patches
corresponding to two states at the late time. Doing so one finds
 \bea
\delta I_{\mathcal{M}} &=& I_{\mathcal{M}} \left[\text{WDW}\vert_{t+\delta t}\right]-I_{\mathcal{M}}\left[\text{WDW}\vert_{t}\right] \nonumber \\
&=& \frac{1}{16 \pi G} \int_{r \to 0}^{r_{h}}\int_{t}^{t+\delta t}\int_{S^{d}}  \; \mathcal{L}_{\rm CG}
  \; d\Omega_{d}\, dt\, dr  \; \ \ \ \text{(at late time)} \nn
&=&\frac{\Omega_{d}\,\delta t}{16 \pi G}\int_{r \to 0}^{r_{h}}r^{d}\left(-2\frac{d+1}{\ell^2}\right)
\left(1-\frac{d^{2}}{2\ell^{2}m^2}\right) dr
=-\frac{\Omega_{d}r_{h}^{d+1}}{8\pi G\ell^{2}} \left(1-\frac{d^{2}
}{2\ell^{2}m^2}\right)\delta t.
\eea
In addition, the contribution of the generalized Gibbons-Hawking terms are
\bea
\delta I_{\partial\mathcal{M}} &=& \frac{1}{16 \pi G} \int_{t}^{t+\delta t} \int_{S^{d}} dt\, d
\Omega_{d} \;\left[-2 \sqrt{f} \; \partial_{r}\left(r^{d}\sqrt{f}\right)\left(1-\frac{d^{2}}
{2\ell^{2}m^2}\right)\right] \Big{\vert}_{r_{h}} \nn
&&-\frac{1}{16 \pi G} \int_{t}^{t+\delta t} \int_{S^{d}} dt\, d\Omega_{d} \;\;\left[-2 \sqrt{f} \;
\partial_{r}\left(r^{d}\sqrt{f}\right)\left(1-\frac{d^{2}}{2\ell^{2}m^2}\right)\right] \Big{\vert}_{r \to 0} \nn
&=&\frac{\Omega_{d}}{8\pi G}\left(d\,
r_{h}^{d-1}+\frac{(d+1)r_{h}^{d+1}}{\ell^{2}}\right)\left(1-\frac{d^{2}}{2\ell^{2}m^2}\right)
\delta t.
\eea	
Here we have considered the contribution of the generalized Gibbons-Hawking terms
 at the horizon $r_{h}$ and the singularity $r \to 0$. Taking bulk and boundary contributions into
 account one arrives at
 \be\label{rateCG}
{\dot{\cal C}}=\frac{d I}{dt}\equiv\frac{d I_{\mathcal{M}}}{dt}+\frac{d I_{\partial\mathcal{M}}}{dt}=2M
\left(1-\frac{d^{2}}{2\ell^{2}m^2}\right)=
2M_{\rm HD}\, .
\ee
Here  to write the final expression we have used the relation between the parameter $M$ and
the physical mass $M_{\rm HD}$ of the critical gravity\cite{Lu:2011zk}. Interestingly enough
this is also consistent with the general expectation of \cite{Brown:2015lvg}.

Note that at the critical point, $m^2=\frac{d^2}{2\ell^2}$ where the model
develops a log gravity, the rate of growth vanishes\footnote{ In four dimensional
critical gravity the action is proportional to the Bach tensor and therefore vanishes for
any Einstein solution\cite{Anastasiou:2016jix}.}.  Actually at this point one has to redo
the computation from the first step with a logarithmic solution that is a time-dependent solution.


\subsection{ Localized Shock wave solutions}
In this subsection in order to arrive at a deeper understanding of the effects of higher  derivatives terms in the growth 
of complexity we will study complexity growth in a black hole perturbed by a shock wave. 
To proceed, following \cite{Brown:2015lvg} we will consider the evaluation of thermofield 
double state perturbed by a single precursor as follows
\bea\label{precursorTFD}
e^{-iH_{L}t_{L}}e^{-iH_{R}t_{R}}W_{x}(t_{w})\mid \text{TFD}\hspace{.5mm}\rangle.
\eea
$H_L$ and $H_R$ are Hamiltonians of the left and right boundary, respectively and 
\bea
W_{x}(t_{w}) =e^{-iHt_{w}}W_{x}e^{iHt_{w}},
\eea
with $W_x$ being  a simple operator  acting on the left boundary.
For  $t_{w}=0$, the precursor operator is a local operator  in the thermal scale,
though it becomes nonlocal by increasing the time $|t_{w}|>0$. In our notations 
the shock  wave emerges at the left boundary at 
$t_w<0$ and we are interested in the region where $t_L,t_R>0$. The geometry dual to the state (\ref{precursorTFD}) can be described by a shock wave solution 
whose metric in the Kruskal coordinates may be given as follows\cite{Roberts:2014isa}
\bea\label{Kruskal.metric}
&&ds^{2}= -2 A(uv)dudv+B(uv)d\Omega^{2}_{d}+2A(uv)h(x)\delta(u)du^{2},\cr &&\cr
&&A(uv)=-\frac{2}{uv}\;\frac{f(r)}{f^2(r_H)},\;\;\;\;\;\;\;\;\;\;B(uv)=r^2,
\eea
where $f(r)$ is given in \eqref{metric} and the Kruskal coordinates in terms of the  
 Schwarzschild coordinates are given by
\bea\label{Kruskalcoord}
u= \exp[\frac{2\pi}{\beta}(r_{\ast}(r)-t)],\hspace{.5cm}v=-\exp[\frac{2\pi}{\beta}(r_{\ast}(r)+t)],
\hspace{.5cm}\beta= \frac{4\pi\ell^{2}}{(d+1)r_{h}+(d-1)\frac{\ell^{2}}{r_{h}}},
\eea
with  $dr_{\ast}=dr/f(r)$. The shock wave strength $h(x)$ may be found from the equations of motion
when the back reaction of the local operator $W_x$ is also taken into account. For Einstein gravity
the shock wave was  studied  in \cite{Roberts:2014isa}. The Kruskal diagram of the geometry 
(\ref{Kruskal.metric}) is shown in Fig.\ref{Kruskal} (see Fig 8 of  \cite{Brown:2015lvg}).
\begin{figure}
	\begin{center}
		\centering
		\vspace{1cm}
		\includegraphics[height=7cm, width=7.5cm]{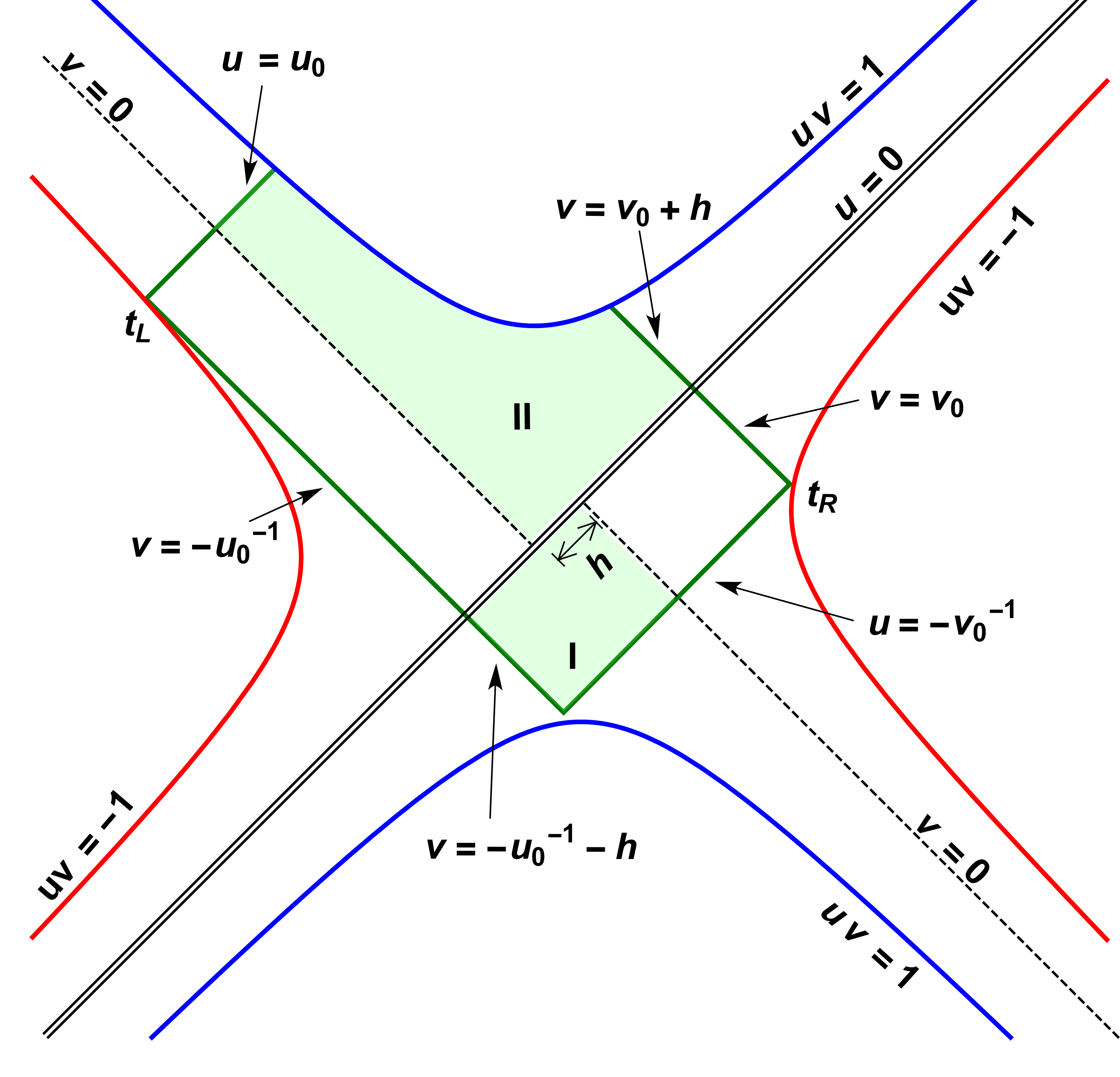}
		\caption{Kruskal diagram of one-shock geometries. The double black line along $u = 0$ is the shock wave. The green lines show the boundary of
		WDW patch. The green shaded region is the intersection of WDW and the black hole 
		interior. In this figure a small shift $h(x)$ with
		$h(x)< v_{0}$ is drown. For a large shift $h(x)$ with $h(x)>v_{0}$ the boundary of WDW
		intersect the past singularity $uv=1$. This figure is drown from Fig 8 of 
		\cite{Brown:2015lvg}.} \label{Kruskal}
		\vspace{0.5cm}
	\end{center}
\end{figure}
As it is clear from the figure, one recognizes two distinctive cases depending on 
the size of the shift $h(x)$. Actually for a small shift where $u_{0}^{-1}+h(x) < v_{0}$ the boundary 
of WDW does not intersect the past singularity leading to a diamond shape (  
region $I$  in the Fig \ref{Kruskal}). On the other hand for large shift where 
$u_{0}^{-1}+h(x) \geq v_{0}$ the past singularity is also intersected by the WDW patch leading 
to a five-sided region similar to region $II$  drown in Fig \ref{Kruskal}. In the latter case  both regions 
above and below $u=0$ axis are five-sided regions bounded by two edges of WDW patch, two horizons and a singularity. As it was argued in   \cite{Brown:2015lvg} due to the fact 
that we are interested in large $t_{R} > 0$, where  $v_{0}^{-1}$ is exponentially small,
the contribution to the action comes from five-sided regions and that of diamond vanishes.
It is worth mentioning that in the Kruskal coordinates one has 
\bea\label{u0v0}
u_{0}=\exp[\frac{2\pi}{\beta}t_{L}],\hspace{1cm}v_{0}=\exp[\frac{2\pi}{\beta}t_{R}].
\eea

Let us now consider the growth of complexity for a shock wave in the critical gravity. To proceed 
we need to find a shock wave solution in a black hole solution of the  critical gravity. 
We note however that following \cite{Brown:2015lvg} in order to  avoid difficulties of having  a spatially compact boundary, one 
may consider the planar-AdS black hole geometry\footnote{ For the compact case, the shock collides with 
itself after turning around the sphere. Nonetheless  for small displacements in the transverse 
direction, the shock wave may be approximated by the plane wave.}.

In critical gravity the planer shock wave has been studied in \cite{Alishahiha:2016cjk} 
where it was shown that due to higher derivative terms there are two butterfly velocities and
\bea\label{h}
h(x) \sim e^{\frac{2\pi}{\beta} (\mid t_{w}\mid-t_{\ast}-\frac{\mid x\mid}{v_B^{(1)}})}-\frac{v_B^{(2)}}
{v_B^{(1)}}e^{\frac{2\pi}{\beta} (\mid t_{w}\mid-t_{\ast}-\frac{\mid x\mid}{v_B^{(2)}})},
\eea 
where $\lambda_{L}=\frac{2\pi}{\beta}$ is the Lyapunov exponent, $t_{\ast}$ is the scrambling time,
and the butterfly velocities are given by
\bea\label{vb}
v_B^{(1)} = \sqrt{\frac{d+1}{2d}},\hspace{1cm} v_B^{(2)} = \sqrt{\frac{d+1}{2d}} \frac{1}{\sqrt{1+
\frac{2\ell^{2}}{d(d+1)} M^{2}}},
\eea
with  $M^{2}\hspace{-.5mm}= \frac{2m^{2}\ell^{2}-d^{2}}{2\ell^{2}}$ being the mass of spin-2 
massive mode.  From these expression  it is clear that $v_B^{(1)} > v_B^{(2)}$.

In what follows we would like to compute the on-shell action of the critical gravity for a shock 
wave solution whose shift $h(x)$ is given by \eqref{h}. To do so we will consider both small and
large shift cases. 

As we have already mentioned for small shift the WDW patch consists of two parts: one five-sided 
and a diamond.  Since ince $v_{0}^{-1} 
= \exp[-\frac{2\pi}{\beta}t_{R}] \ll 1$, the diamond region is exponentially small and has a vanishing 
contribution to the action and therefore one only needs to compute the contribution of the five-sided 
region.  The corresponding contribution contains both the bulk and boundary parts.
Taking into account that we are in the small shift region where  $h(x) < v_{0}$ one  finds  
\bea\label{smallshock1}
I_{\mathcal{M}} = -\frac{L^{d}r_{h}^{d+1}}{8\pi G\ell^{2}} \left(1-\frac{d^{2}}{2\ell^{2}m^{2}}\right)
\frac{\beta}{2\pi}\log{u_{0}v_{0}},
\eea
and
\bea\label{smallshock2}
I_{\partial\mathcal{M}} =-\left[\frac{L^{d}}{8\pi G}\left(d\,r^{d-1}+\frac{(d+1)r^{d+1}}{\ell^{2}}
\right)\left(1-\frac{d^{2}}{2\ell^{2}m^{2}}\right)\right]^{r_{h}}_{0}\frac{\beta}{2\pi}\log{u_{0}v_{0}}.
\eea
Note that the generalized Gibbons-Hawking terms are computed at the singularity and at
the horizon. Here $L^{d}$ is spatial volume of the internal space.  Putting everything together one 
arrives at 
\be
I = 2M_{\text{HD}}(t_{L}+t_{R}).
\ee
To get this expression we have used the equation (\ref{u0v0}) and the 
definition of physical mass in critical gravity. It is exactly the same form we have 
for the Einstein gravity except with inserting the proper mass of the model. 
This expression might be understood from the fact that at small shift the explicit form of $h(x)$ does 
not play an essential role. In other words it means that for large $t_{R}$, we have evolved the state 
to a region  where the back-reaction of perturbation (precursor operator) is negligible and 
the complexity is simply given by time evolution of  TFD state.

Let us now consider the big shift where $h(x)>v_0$. In this case we have to deal with 
a WDW patch 
consisting of two parts in the shape of the five-sided region  and therefore both parts 
contribute to the on-shell action.  The  contribution of the upper five-sided region 
(taking onto account   bulk and boundary terms) is
\bea\label{bigShockup1}
&& I_{\text{upper}} = \frac{L^{d-1}}{8\pi G}\left(d r_{h}^{d-1}+\frac{dr_{h}^{d+1}}{\ell^{2}}\right)\left(1-
\frac{d^{2}}{2\ell^{2}m^{2}}\right)  \frac{\beta}{2\pi}\int dx\hspace{1mm} \log [u_{0}h(x)],
\eea
while that of lower one is
\bea\label{bigShockup2}
&& I_{\text{lower}} = \frac{L^{d-1}}{8\pi G}\left(d r_{h}^{d-1}+\frac{dr_{h}^{d+1}}{\ell^{2}}\right)\left(1-
\frac{d^{2}}{2\ell^{2}m^{2}}\right) \frac{\beta}{2\pi}\int dx\hspace{1mm} \log[v_{0}^{-1}h(x)].
\eea
To find the above expression we have used that fact at $h(x)>v_0$ and $u_0^{-1}\ll 1$.
Moreover the overall factor $L^{d-1}$ in both expressions comes from the fact that the 
corresponding  shock wave \eqref{h} is localized in one transverse direction. From 
equations (\ref{bigShockup1}) and(\ref{bigShockup2}) one arrives at
\bea\label{complexitylocalsch1}
I= \frac{L^{d-1}}{8\pi G}\left(d r_{h}^{d-1}+\frac{dr_{h}^{d+1}}{\ell^{2}}\right)\left(1-\frac{d^{2}}
{2\ell^{2}m^{2}}\right)  \frac{\beta}{2\pi}\int dx\hspace{1mm} \log [u_{0}v_{0}^{-1}h^{2}(x)].
\eea
On the other hand by making use of equations (\ref{h}) and (\ref{u0v0}), the integrand part
of the above equation may be simplified as follows
\be \label{integrandShock}
\frac{\beta}{2\pi}\log[u_{0}v_{0}^{-1}h^{2}(x)]=(t_{L}+t_{R})+2\left(|t_{w}|
-t_{\ast}-t_{R}-\frac{|x|}{v_{B}^{(1)}}\right)
+2\hspace{1mm}\frac{\beta}{2\pi}\log\left[1-\frac{v_{B}^{(2)}}{v_{B}^{(1)}}e^{-
\frac{2\pi}{\beta} (\frac{1}{v_{B}^{(2)}}-\frac{1}{v_{B}^{(1)}}) |x|}\right].
\ee
It is now easy to perform the integration over $x$ to get the following expression for 
the on-shell action\footnote{{It is worthwhile to mention that for spherically symmetric shock
wave emerging from the boundary at time $t_{w}$, instead of (\ref{h}) we have $h \sim e^{\frac{2\pi}{\beta}(\mid\hspace{-.2mm}t_{w}\hspace{-.2mm}\mid-t_{\ast})}$ which implies that $I=2M_{\text{HD}}\big[t_{L}-t_{R}+2(\mid\hspace{-.5mm}t_{w}\hspace{-.5mm}\mid-t_{\ast})\big].$}}
\bea\label{IShockkk}
I& =& 2M_{\text{HD}} \left(t_{L}+t_{R}\right)+4\tilde{M}L^{d-1}\frac{v_{B}^{(1)}}{2}\bigg(\mid
\hspace{-.5mm}t_{w}\hspace{-.5mm}\mid-t_{\ast}-t_{R}\bigg)^{2}\cr \nonumber\\
&&-4\tilde{M}L^{d-1}\frac{\beta^{2}}{4\pi^{2}} \frac{v_{B}^{(1)}v_{B}^{(2)}}{(v_{B}^{(1)}-v_{B}^{(2)})}
\left[{\rm Li}_{2}\left(\frac{v_{B}^{(2)}}{v_{B}^{(1)}}\right)-{\rm Li}_{2}
\left(\frac{v_{B}^{(2)}}{v_{B}^{(1)}}
e^{\frac{-2\pi (v_{B}^{(1)}-v_{B}^{(2)})(\mid\hspace{-.2mm}t_{w}\hspace{-.2mm}
\mid-t_{\ast}-t_{R})}{v_{B}^{(2)}\beta}}\right)
\right],
\eea
where $\tilde{M}$ is the spatial energy density of the corresponding black hole
and ${\rm Li}_2(x)$ is the dilogarithm function. 
It is worth noting that to find the above expression we have used the fact that 
we are in the big shift approximation  where $h(x)>v_0$ we have
\bea\label{timeintegral}
e^{\frac{2\pi}{\beta}\big(\mid t_{w}\mid -t_{\ast}-t_{R}\big)}\left(1-\frac{v_{B}^{(2)}}{v_{B}^{(1)}}e^{-
\frac{2\pi}{\beta} \left(\frac{1}{v_{B}^{(2)}}-
\frac{1}{v_{B}^{(1)}}\right)\mid x\mid}\right)>1,
\eea
and the integration over direction $x$ has been performed in the range of
$ 0\leq x\leq v_{B}^{(1)}(|t_{w}| -t_{\ast}-t_{R})$. Actually one observes that the first two
terms in equation \eqref{IShockkk} are the same as those in Einstein  gravity
\cite{Brown:2015lvg}, though the last term is the effect of higher derivative terms that 
slows down the rate of complexity growth.


\section{Discussions}

In this paper using ``complexity=action'' proposal, we have studied complexity of 
black hole solutions in some certain gravitational theories. In this proposal the complexity 
of a holographic state 
is given by the on-shell action evaluated on the Wheeler-De Witt patch.
The main challenge to evaluate the on-shell action is to find the contribution of boundary terms defined on different time like or null boundaries  or on intersection of these boundaries.
 
Although for the Einstein gravity these null boundary terms have been recently obtained
in \cite{Lehner:2016vdi}, for a general higher derivative action such boundary terms are not 
known. Indeed it is worth mentioning that even the standard Gibbons-Hawking terms
for general action have not fully understood yet. In order to circumvent this issue, following 
 \cite{Brown:2015lvg} we will compute the rate of complexity growth in which the generalized 
 Gibbons-Hawking boundary terms are enough. Of course we have considered certain  
 gravitational theories with higher order terms whose generalized Gibbons-Hawking terms are 
 known. These include  $F(R)$-gravity and critical gravity in diverse dimensions. 
  Encouraged by the results of \cite{Chapman:2016hwi} we would expect that the final results 
 are independent of the missing null boundary terms of our models.
 
We have shown that for the neutral black hole solution the complexity growth saturates 
the proposed bound when the results are written in terms of physical quantities of the model 
under consideration
\be
{\dot {\cal C}}=2M,
\ee
where $M$ is the physical mass (ADM mass ) of the corresponding black hole
(see also \cite{Pan:2016ecg} for massive gravity)\footnote{This is also consistent 
with the result of \cite{Yang:2016awy} where it was shown that under certain conditions
uncharged black holes  have the fastest computational complexity growth.}. 
It was important to note that 
in higher derivative gravities the parameter that explicitly appears in the  solution is not necessarily 
the physical mass of the solution.  

We have also studied  the effect of a single  shock wave in the growth of complexity in the critical 
gravity. Due to the presence of higher derivative terms in the action the shock wave in the critical
gravity propagates with two different butterfly velocities. We have seen that the complexity 
growth are affected by both velocities. Although the effect of massless mode is similar to that of
Einstein gravity the effect of massive mode slows down the rate of complexity growth. Since 
we do not have a good understanding of complexity in the field theory dual, it might be 
hard to confirm this behavior from field theory point of view. Therefore our results should be treated 
as a prediction for the holographic complexity.  

We could have also studied complexity growth for a shock wave in  $F(R)$ gravity. 
We note, however,
that since the model has only massless gravity mode, the situation would qualitatively be similar
to that of Einstein gravity. This might be understood  from the fact that $F(R)$ gravity may be
map to an Einstein gravity coupled to a scaler field. In this case one gets only one butterfly
velocity which  comes from the gravitational part that  could affect the growth.

As a final remark we would like to recall that there is another proposal for 
complexity known as ``complexity=volume'' or in short CV proposal
\cite{{Stanford:2014jda},{MIyaji:2015mia},{Couch:2016exn},{Carmi:2016wjl}}.   
It is also expected \cite{Alishahiha:2015rta}  
that in the presence of higher derivative terms one needs to extend the notion of volume to a new
quantity. The situation is similar to thermal and entanglement entropies where the area must
be replaced by certain functional 
to be evaluated on the horizon and RT curve, respectively (see {\it e.g.} \cite{Dong:2013qoa}). 
As for complexity we would also like to have a new functional to be evaluated on a co-dimension
one hypersurface. Although it is not clear to us whether the CV proposal fulfills all requirements 
to be identified with complexity,   since in the cases we have been considering in the 
previous sections  we would expect that CV and CA proposal would qualitatively match,  
in what follows we shall suggest two possibilities that are consistent with our 
results  presented in the previous sections.

Motivated by Wald formula for entropy one may consider an expression which has a Wald-like form.
To be precise consider a co-dimension one time slice whose normal vector is given by
$n_\mu^{1}$. Then let us foliate the co-dimension one hypersurface alone the radial coordinate
parametrized by the normal vector $n_{\mu}^2$. Having two normal vector one may
construct  a binormal anti-symmetric 2-tensor by which we can compute  Wald integrand as follows
\bea
\frac{\partial \mathcal{L}}{\partial  R_{\mu\nu\alpha\beta}}\epsilon_{\mu\nu}\epsilon_{\alpha\beta}.
\eea
Now the natural proposal for CV complexity is\cite{Alishahiha:2015rta}
\be\label{CV1}
C_V=-\frac{1}{\ell}\int_{\cal B} \frac{\partial \mathcal{L}}{\partial  R_{\mu\nu\alpha\beta}}\epsilon_{\mu
\nu}\epsilon_{\alpha\beta},
\ee
where $\ell$ is a length scale in the bulk gravity and ${\cal B}$ is a co-dimension one maximal  
time slice. Using the explicit expressions of Ricci and scalar curvature for  Black hole solutions in the
critical gravity \eqref{metric} one gets
\be
\frac{\partial \mathcal{L}_{\text{HD}}}{\partial  R_{\mu\nu\alpha\beta}}\epsilon_{\mu\nu}
\epsilon_{\alpha\beta}= -2\left(1-\frac{d^{2}}{2\ell^{2}m^2}\right).
\ee
Plugging this expression in the equation \eqref{CV1} one finds the CV complexity for the critical
gravity which is consistent with our results found in the previous section.

On the other hand in the context of the entanglement  equilibrium the authors of
 \cite{Bueno:2016gnv} has found an expression for volume that  kept fixed while varying 
 the entanglement entropy. Based on this result the authors  have proposed an expression for CV complexity which is given as
follows. Let us denote the normal vector to the time slice hypersurface by $n^\mu$ and define
the following functional for given Lagrangian
\be
\frac{\partial \mathcal{L}}{\partial  R_{\mu\nu\alpha\beta}}\left(a\hspace{1mm} n_{\mu}n_{\beta}h_{\nu
\alpha}+b\hspace{1mm} h_{\mu\beta}h_{\nu\alpha}\right)+c,
\ee
where $a,\,b$ and $c$ are three arbitrary constants.  Then the generalized CV complexity may be
defined as follows
\be\label{CV2}
C_V=-\frac{1}{\ell}\int_{\cal B} \left[\frac{\partial \mathcal{L}}{\partial  R_{\mu\nu\alpha\beta}}\left(a\hspace{1mm} n_{\mu}n_{\beta}h_{\nu
\alpha}+b\hspace{1mm} h_{\mu\beta}h_{\nu\alpha}\right)+c\right],
\ee
where as before  $\ell$ is a length scale in the bulk gravity and ${\cal B}$ is a
co-dimension one maximal time slice. Since the complexity in terms of volume may be defined
up to an overall factor of order one, we may set $a=1$. It is an easy excesses
to evaluate the above functional for
the black hole solution \eqref{metric} in critical gravity. Doing so one arrives at
\be
\frac{\partial \mathcal{L}_{\text{HD}}}{\partial  R_{\mu\nu\alpha\beta}}\left(\hspace{1mm} n_{\mu}
n_{\beta}h_{\nu\alpha}+b\hspace{1mm} h_{\mu\beta}h_{\nu\alpha}\right)+c= \frac{d+1}{2}(1-d\,b)
\left(1-\frac{d^{2}}{2\ell^{2}m^2}\right)+c.
\ee
It is then clear that to get a consistent result one should set $c=0$ and $b=\frac{d+5}{d(d+1)}$.
Since in this definition the generalized volume is ambiguous up to an overall factor of order one, different $a$ and $b$ may also be chosen, though it seems that the constant $c$ should 
always be set to zero.

Definitely it would be interesting to further explore properties of the above proposals 
for the generalized volume and to their connection with holographic complexity.

\section*{Acknowledgments}

We would like to thank Amin Akhavan, Ali Mollabashi,  Mohammad M. Mozaffar, Dan  A. Roberts,
Ahmad Shirzad and Mohammad R. Tanhayi for useful discussions.
This work is supported by Iran National Science Foundation (INSF). 
We would also like to thank the referee for his/her useful comments.

\end{document}